\documentclass[11pt,a4paper]{article}
\usepackage{acl2015}
\usepackage{times}
\usepackage{url}
\usepackage{latexsym}

\usepackage{graphicx}
\usepackage[fleqn]{amsmath}
\usepackage{algorithm}
\usepackage{algpseudocode}
\usepackage{booktabs}
\usepackage{tabularx}

\DeclareMathOperator*{\argmin}{\arg\!\min}

\title{Query Clustering using Segment Specific Context Embeddings}

\author{S.K. Kolluru \\
  IIT Bhubaneswar \thanks{\hspace{0.1cm} work done as part of internship at Microsoft, India}  \\
  {\tt kk13@iitbbs.ac.in} \\\And
  Prasenjit Mukherjee\\
  Microsoft, India\\
  {\tt pmukherj@microsoft.com} \\}

\date{}

\begin{document}
\maketitle
\begin{abstract}
	
This paper presents a novel query clustering approach to capture the broad
interest areas of users querying search engines. We make use
of recent advances in NLP - word2vec and extend it to get
query2vec, vector representations of queries, based on query
contexts, obtained from the top search results for the query 
and use a highly scalable Divide \& Merge 
clustering algorithm on top of the query vectors, to get the clusters.
We have tried this approach on a variety of segments, including Retail,
Travel, Health, Phones and found the clusters to be effective in
discovering user's interest areas which have high monetization potential.

\end{abstract}

\section{Introduction}

According to the February '16 comScore reports \footnote{https://www.comscore.com/Insights/Rankings/comScore-Releases-February-2016-US-Desktop-Search-Engine-Rankings}, the number of core searches in desktops exceeds 16 billion in the month of February alone. With the present scale of searches, manual identification of areas people query for, is far from possible. But with the current trend in modern search engines to provide tailor-made experiences in specific domains people query for, the need to identify user's interest areas becomes more imperative. This trend benefits both the customers, who gain from the domain-specific information search engines provide and the search engines, in turn, gain more popularity leading to higher ad revenues. In this current scenario, automatic identification of these promising areas which people are interested in (expressed through their queries made to search engines), and which also have a high revenue share becomes an important task. This allows search engines to focus their resources in providing better experiences to queries falling in these identified areas. Along with the issue of scalability, manual identification also suffers from chances of missing less-known, location or time specific areas. In an attempt to automatically identify these areas, the paper presents a novel unsupervised query clustering approach with specific emphasis on scalability so that the approach can be of practical utility, considering the current scale of search scenario.
 
An important problem in effective query clustering is the computation of similarity between queries. Inspired by the recent advances in NLP, namely word embedding generation using word2vec (\newcite{mikolov2013distributed}), where similarity between words has been effectively computed using cosine similarity between embeddings of words, generated by passing a large text corpus to either the Continuous Bag of Words(CBOW) or Skip-Gram neural network architectures, we propose to solve the problem of query similarity by generating embeddings for each query, based on search engine results for the query. We start with the premise that, the top results given by search engines are often relevant to the queries. So we can use those results themselves, to generate meaningful embeddings for the queries. In major search engines (Bing and Google), the results for queries contain the url of the linked page, the title and 3-4 line summary of what is in the page that is relevant to the query, which are called snippets. These information convey a lot about the true meaning of the query. So, we consider them as the context of the query, and augment word2vec, with an additional step to generate embeddings of the query from it's context. Thus we generate context for the query from the search engine results and use the context to get embedding for the query. In this way arrive at query2vec which generates embeddings from queries. We train our query2vec model on a segment specific corpus for best results. 

The main reason for choosing to generate embeddings for queries from it's context is the observation that similar queries will have similar contexts, thus giving similar queries, similar embeddings. The embedding derived in this manner, exhibit interesting properties, with `apple iphone' and `samsung galaxy' having embeddings with a high cosine similarity when compared to their cosine similarity with `dell laptops'. Interesting query-analogies have also surfaced using these embeddings, such as, ``apple 6s'' - ``apple'' + ``samsung'' = ``samsung galaxy'' ! The embeddings were sufficiently efficient to capture that the relation between ``apple 6s'' and ``apple'' was a phone-to-company relation and ``samsung'' holds the same relation with ``samsung galaxy''.

In this way, we capture deeper semantic similarity of queries than recent query clustering approaches, \newcite{hong2016accurate}, also computes query similarity using top-k search engine results, but only considers the common URLs, as they are based on the intuition that similar queries will have common URLs showing up in the search results. They verify their hypothesis on queries such as ``Honda accord Toyota camry" and ``Civic vs. Corolla" which have common URLs in search engine results. But we found that the same doesn't hold for the queries ``Civic" and ``Corolla" which surface no common URLs in the top 10 search results  of Bing but do share certain keywords in the top results such as `cars',`reviews',`sedan',`automobile',`prices' which our method leverages by taking them as the context for query and then generating embeddings for the query.

With well over 15 billion searches being done every month, the problem of scalability lies at the heart of query clustering. Query clustering based on agglomerative clustering or similar techniques are bound to fail due to the shear scale of the data they have to operate on. Their alternative, K-Means, though light-weight suffers from the requirement to know the `K', expected number of clusters. To address both of the above problems, we implement a Divide \& Merge Clustering approach, which does not require knowing the initial number of clusters and at the same time, akin to K-Means, is light-weight. We show in the results, Section \ref{Results}, how the combination of query context embeddings and Divide \& Merge Clustering has proven effective in clustering queries of various segments. For example, some of the interesting clusters formed on clustering queries of the retail segment include `cyber monday', `black friday', `gift cards'. These areas are known high-revenue earners for search engines, thus demonstrating the potential of our method to automatically identify user's interest areas.

The remainder of the paper is organized as follows. We discuss the related work in Section \ref{Related Work}, present the technique used for query embedding generation in Section \ref{QueryEmbedding}, detail the Divide \& Merge Clustering in Section \ref{Clustering}, show the experimental results on queries of Bing search engine in Section \ref{Results} and finally conclude in Section \ref{Conclusion}.



\section{Related Work} \label{Related Work}

The major problem of Query Clustering is the short and ambiguous nature of queries. Therefore, Jaccard similarity of queries based on common keywords suffers from the sparse nature of queries. To address this problem, \newcite{beeferman2000agglomerative} used an agglemorative clustering algorithm where similar queries are discovered by treating the queries and the clicked-URLs as a bipartite graph. \newcite{wen2002query} used click-through information, along with the query keywords and used DBSCAN for clustering. In \newcite{chuang2002towards}, they construct a feature bag for every query based on the results returned by search engine for that query and then generate TF-IDF vectors for the query from the words in it's feature bags, followed finally by Agglemorative Clustering to generate a query taxonomy. In \newcite{meng2013new} they calculate query similarity using feature words picked up from user-clicked documents (for the query) and calculating similarity between feature words based on WordNet (\newcite{fellbaum1998wordnet}), followed by Agglemorative Clustering. In \newcite{uddin2015top}, the top-k URLs, text similarity along with time similarity are considered for calculating query similarity and clustering is performed using SOM.

The most recent methods in query clustering \newcite{hong2016accurate}, calculates query similarity based on the top-k URLs returned by a search engine for that query, termed as transitional similarity in their work. They rely on the fact that URLs given by search engine results are a good indication of query similarities. Since we use the context of the query and generate embeddings for the query from the context, we consider deeper semantic similarity of queries than just using common URLs.

Even the most recent works don't make use of the latest advances in word embedding generation for improving query similarity measurement. For example, adapting the word2vec architecture presented in \newcite{mikolov2013distributed} for queries, by training the queries to predict the clicked documents (as in \newcite{huang2013learning},\newcite{mitra2016dual}), similar queries get mapped close to each other. In \newcite{yang2015learning}, query-vectors are generated by training queries to predict the topic distribution and clicked URLs for the purpose of query classification.

Our work can be seen as bridging the above identified gap by using query similarity based on embeddings, for the goal of query clustering. 

In \newcite{metzler2007similarity}, expanded representation of query is formed by taking the titles and snippets of the results of search engine. Similar to ours, the expanded representation is used for computing query similarity, with the difference being our use of word2vec. For clustering without any prior assumptions on the structure of data (such as knowing the number of clusters), we use the idea from \newcite{cheng2006divide}, \newcite{fred2002data}, where they have a divide phase followed by merge phase to generate clusters.

\section{Query Embedding Generation} \label{QueryEmbedding}

For query clustering, we require a suitable mathematical representation of a query which captures the semantic similarity of queries. For example, `apple iphone' should appear closer to `samsung galaxy' than to `dell laptop' in such a representation, which matches our common knowledge that `apple iphone' and `samsung galaxy' are more related to each other, both being smart phones, than they are to `dell laptop'.

Such representations of words, which capture the semantic similarity, have been extensively reported by the NLP community, \newcite{rumelhart1988learning}, \newcite{bengio2003neural}. It has recently gained traction, due to high quality representations produced by \newcite{mikolov2013distributed}, \newcite{mikolov2013efficient}, using Skip-Gram and CBOW architectures. Before we proceed to show how we generate such vector representations for queries, it would be worthwhile to review the methodology used in word2vec, as we augment it to generate vector representations for queries.

\subsection{CBOW and Skip-Gram Architectures} \label{word2vec}

The basic assumption for generating word vectors is that, similar words appear in similar contexts. In order to generate suitable vector representations of words, which capture the semantic similarity between words, 2 neural network architectures have been proposed  - Continuous Bag of Words (CBOW) and Skip Gram. In order to make the vector generation computationally feasible, Negative Sampling is used as the objective function.

The input to the system is a text corpus. A window of fixed size is slided along the text corpus, where the word in the middle is treated as the target word and remaining words in the window are treated as context words. These are given as input or expected output to the neural net, depending on whether the method used is CBOW or Skip-Gram. In CBOW, the word embeddings are trained to predict the target word given the context words, while in Skip-Gram the word embeddings are trained to predict the context words around that target word. In this way, the supervised neural network is trained in a surprisingly unsupervised manner.  The negative sampling approach is used to reduce the computational complexity of both the methods by limiting the number of word  vectors updated for each training sample. The number of updated word vectors would have been equal to the size of vocabulary if softmax would have been used as the objective function. Instead in negative sampling, only a fixed number of negative-samples are updated (independent of vocabulary size). The sampling can be done assuming some noise distribution (such as unigram).

\subsection{Query Context} \label{QueryContext}

To generate embeddings for a query, we first derive the context of the query. The context of the query is generated by concatenating the title, snippets and URLs of top-10 search engine results for the query. 

\begin{figure}[ht!]
\centering
\includegraphics[width=80mm]{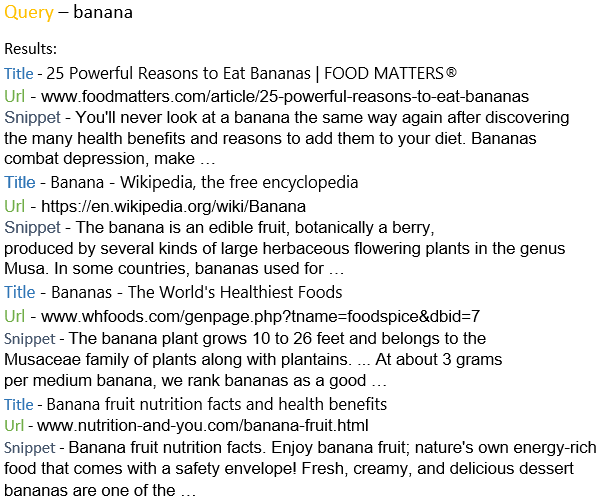}
\caption{Query Context of `banana' \label{QueryContextFig}}
\end{figure}

For example, consider the sample search query `banana' and some results given by a search engine for that query as shown in Figure  \ref{QueryContextFig}. The query context for `banana' would be the concatenation of titles, URLs and the snippets of the top-10 results generated by the search engine (only 4 results shown in figure for brevity). The URLs are broken into tokens and words such as http, https, www, com, etc are removed and the remaining tokens are added to the query context. Furthermore, stopword removal is performed on the query context to remove common words which do not contribute to the meaning of the query such as to, the, a, for, etc. To represent this symbolically,
\begin{equation}
\begin{split}
QueryContext(query) = \bigcup_{i=1}^{10}  \{title(queryResult_i) \\ 
                                \cup \: url(queryResult_i) \: \cup \: snippet(queryResult_i)\}
\end{split}
\end{equation}

\subsection{Query2Vec} \label{Query2Vec}

After generating query contexts for all the queries in the above-mentioned way, we generate vectors (the terms, embeddings and vectors are used interchangeably with one other) for the queries using the following 2-step process:

\begin{itemize}
  \item The query contexts of all queries are concatenated to form a text corpus, from which word embeddings are generated through word2vec.
  \item The vector for a query is generated by computing the normalized sum of vectors of words in it's context.
\end{itemize}

If we denote the set of all queries as $Q$ and set of query contexts as $C$, where $C$ is generated in the following manner,
\begin{equation}
C = \{ c \mid c = QueryContext(q) \; \forall \: q \: in \: Q\}
\end{equation}

A text corpus is generated by concatenating all the query contexts in $C$. If the set of queries, $Q$ belong to a specific segments, such as retail, health or travel, then the corpus we get is a segment-specific corpus, since all contexts in $C$ belong the that segment. So we denote the segment specific corpus as $SSCorpus$. We train a word2vec model using $SSCorpus$. In the Results Section, we show how training the word2vec model on the segment specific corpus produces significantly better clusters. 
\begin{equation}
SSCorpus = \bigcup \: c \: , \: \qquad \forall \: c \in \: C
\end{equation}

The skip-gram method is used to train the word2vec model with $SSCorpus$ as the input. Let us denote the trained model using the function $Word2Vec$ which takes as input a word from $SSCorpus$ and gives as output the vector generated from the model, i.e,
\begin{equation}
v = Word2Vec(w) \: , \qquad w \in SSCorpus
\end{equation}

For each query, vectors are generated by taking the sum of Word2Vec outputs for every word in it's context, and finally normalizing it, to remove differences arising due to varying number of words in every query's context. We denote this operation using the function $Query2Vec$ which takes as input the query and returns the vector generated for that query.
\begin{equation}
\begin{aligned}
Query2Vec&(query) = \\
 &norm(\sum Word2Vec(w)) \\ 
&\forall \: w \: \in \: QueryContext(query)
\end{aligned}
\end{equation}

The vectors generated in this manner maintain the semantic similarity between queries, measured using cosine similarity. For example, when we applied this to a dataset 100 K queries of retail segment taken from Bing search engine, the 5 nearest queries to `\textbf{kyrie irving shoes}' were found to be `\textbf{nike sb}',`\textbf{nike air mags}',`\textbf{under arnour shoes}',`\textbf{lebron james shoes}',`\textbf{nike jordans}', which all happen to be related to shoes, demonstrating the quality of query-vectors generated in this fashion.

\begin{figure*}[ht!]
\centering
\includegraphics[width=\textwidth,height=7cm,keepaspectratio]{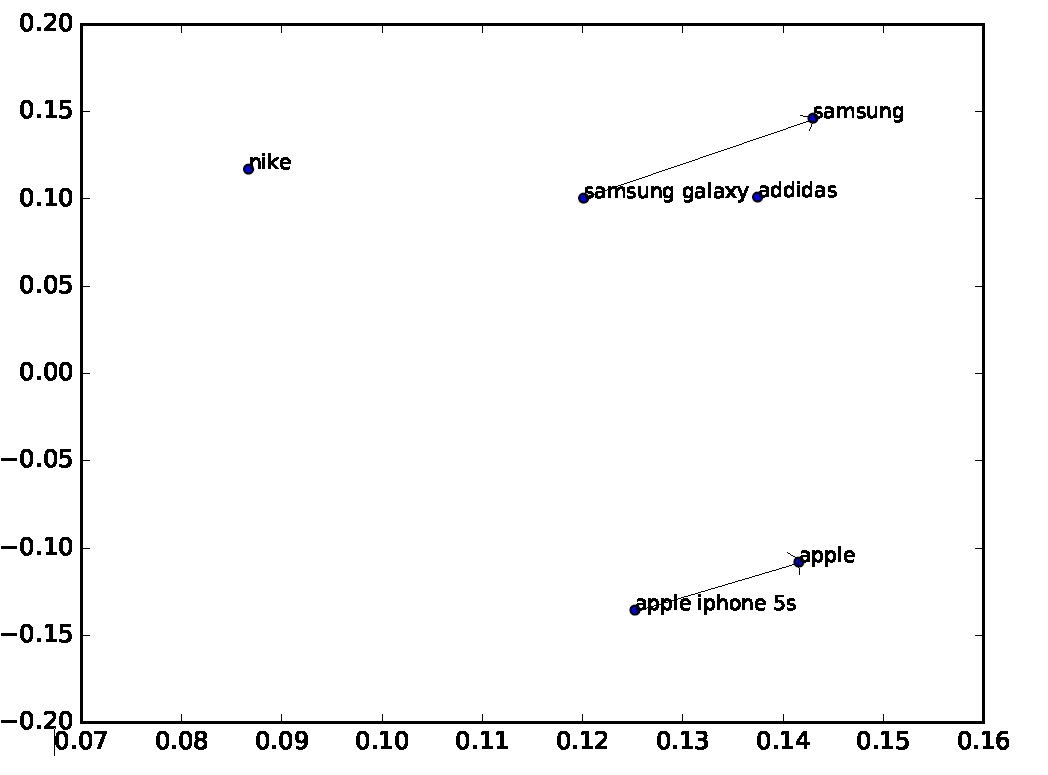}
\caption{Query Analogy - \: ``apple 6s'' - ``apple'' + ``samsung'' = ``samsung galaxy''  \label{AnalogyVizFig}}
\end{figure*}

Even more fascinating is the fact that, similar to word analogies of word2vec, interesting query analogies have also surfaced using this method. For example the query closest to ``apple 6s'' - ``apple'' + ``samsung'' happens to be ``samsung galaxy'', matching with common knowledge that the difference between ``apple 6s'' and ``apple'' is that of a phone and ``samsung'' holds the same relation with ``samsung galaxy''. This is graphically illustrated in Figure \ref{AnalogyVizFig}, where 2 other points - ``addidas'' and ``nike'' are added as control points. From the figure we can see how the vector differences have nearly the same directions, making their cosine similarity very high.

\section{Divide \& Merge Clustering} \label{Clustering}

After the query vectors have been generated, to address the problem of query clustering, taking inspiration from \newcite{cheng2006divide}, \newcite{fred2002data}, we use our own clustering algorithm called, Divide \& Merge Clustering, which does not require prior information about the data (such as number of clusters), but at the same time is computationally efficient (like K-Means). The algorithm requires more number of parameters than K-Means (which is just the K), but they are easily configurable and we have observed that the same set of parameters have worked well for a number of different datasets.

In a nutshell, the method (See Algorithm \ref{DMAlgo}) starts with a single cluster, iteratively splits the clusters (Divide Phase) until convergence is reached and finally to correct any errors that may have occurred in divide phase, merges nearby clusters (measured by distance between cluster centres).

Since, we are dealing with the specific case of search queries, 2 parameters associated with them are their $Impression$ and $Entropy$ values. $Impression$ is the number of times the search engine received the query while the $Entropy$ is the measure of randomness associated with user's click-responses for that query, measured using the following formula (assuming k links are associated with a query) :
\begin{equation}
Entropy = \sum_{i=1}^{k} p_i * log_2(p_i)
\end{equation}
where $p_i$ is the probability of the $i^{th}$ link being clicked. Queries which have high impression and high entropy provide maximum information, as more people are interested in them and they are not just navigational queries (only used for navigating to the website, not for searching something, which typically have low entropy values). So as to give them greater importance, the centre update in Line 9 of Algorithm \ref{DMAlgo} is weighted with $Entropy$ and $Impression$ (logarithm of $Impression$ is taken to remove differences in scale). 

In Algorithm \ref{DMAlgo}, the label $l_i$ refers to the cluster assignment of query $q_i$. The distance metric between 2 vectors is defined as cosine similarity of the vectors subtracted from 1 (See Line 1 of Algorithm \ref{Procedures}). After every divide phase, some iterations are provided for the clusters to stabilize, referred to as $Stabilizing\:Iterations$, during which K-Means algorithm is applied. The next divide phase starts when the $Stabilizing\:Iterations$ are completed (referred to as $DividePhaseCriterion$ being satisfied in line 12 of Algorithm \ref{DMAlgo}). This is checked by measuring the ratio of queries migrating between clusters. If it is below a threshold value ($StabilizingThr$) or the maximum permitted $Stabilizing Iterations$ ($Max\_Stablizing\_Iterations$) are reached, then the next divide phase begins.

The Divide Phase of the algorithm stops when the clusters have converged (if the percentage change in number of clusters is less than a threshold value, $ExitThr$) or $Max\_Iterations$ have been completed. Finally, the Merge Phase of the algorithms executes, merging nearby clusters. 

\begin{algorithm}
  \caption{Divide \& Merge Clustering} \label{DMAlgo} 
  \begin{algorithmic}[1]
    \Require Query Vectors $q_1..q_m$ , Entropies $e_1..e_m$, Impressions $imp_1..imp_m$
    \Ensure Labels $l_1..l_m$
    \State Randomly initialize centre $c_1$
    \State $\mathcal{C} \leftarrow \{c_1\}$
    \State $K \leftarrow 1$
    \Repeat
      \For{$i \leftarrow 1$ to $m$}
        \State $l_i \leftarrow \argmin\limits_{j \leftarrow 1 \: \text{to} \: K} \{\Call{dist}{q_i,c_j}\}$
      \EndFor
      \For{$j \leftarrow 1$ to $K$}
        \State $c_j \leftarrow \sum_{\substack{i=1, \\ l_i==j}}^{M} q_i * e_i * log_2(imp_i)$
	\State $c_j \leftarrow \frac{c_j}{\|c_j\|}$
      \EndFor
      \If{$Divide Phase Criterion$ satisfied}
	 \State $\mathcal{C} \leftarrow \Call{Divide}{\mathcal{C}}$ \Comment Divide Step
	 \If{$\Delta \hspace{0.1cm} len(\mathcal{C}) < ExitThr$ }
	    \State $\textbf{break}$
         \EndIf
	 \State $K \leftarrow len(\mathcal{C})$
      \EndIf
    \Until{$Max\_Iterations$ reached}
    \State
    \State $\mathcal{C} \leftarrow \Call{Merge}{\mathcal{C}}$ \Comment Merge Step
    \State $K \leftarrow len(\mathcal{C})$
    \State
    \For{$i \leftarrow 1$ to $m$} \Comment Recompute Labels
      \State $l_i \leftarrow \argmin\limits_{j \leftarrow 0 \: \text{to} \: K} \{dist(q_i,c_j)\}$
    \EndFor
    \State
    \Return $l_1..l_m$
  \end{algorithmic}
\end{algorithm}

Any criterion can be chosen for division of cluster (see Line 8 of Algorithm \ref{Procedures}). In our implementation, we assume a Gaussian distribution of query vectors from the centre of the cluster. Then we find those queries which are greater than standard deviation away from the centre (the previously defined distance metric is used for measuring distance of query vector from centre) and group them into a cluster of their own. The queries within standard deviation form a cluster of their own. If both of the split clusters have a size greater than the minimum defined size of a cluster (referred to as $MinSizeCriterion$ in Line 9 of Algorithm \ref{Procedures}), then the centres of the split clusters are added into the list of split centres ($\mathcal{SC}$), else the split is undone.

\begin{algorithm}
  \caption{Procedures} \label{Procedures}
  \begin{algorithmic}[1]
    \Procedure{dist}{$v_1,v_2$}
	\State
	\Return 1 - $v_1 \cdotp v_2$
    \EndProcedure
    \State
    \Procedure{Divide}{$\mathcal{C}$}
      \State $\mathcal{SC} = \{\emptyset\}$
      \For{$c \: \text{in} \: \mathcal{C}$}
        \State $\{c_i,c_j\} = split(c)$
	\If{$Min Size Criterion$ satisfied} 
	  \State $\mathcal{SC} = \mathcal{SC} \cup \{c_i,c_j\}$
        \Else
	  \State $\mathcal{SC} = \mathcal{SC} \cup \{c\}$
        \EndIf
      \EndFor
      \State
      \Return $\mathcal{SC}$ 
    \EndProcedure
    \State
    \Procedure{Merge}{$\mathcal{C}$}
      \For{$c_i,c_j \: \text{in} \: \mathcal{C}$}
        \If{$\Call{dist}{c_i,c_j} < MergeThr$}
	  \State $c = \frac{c_i+c_j}{\|c_i+c_j\|}$
	  \State $\mathcal{C} = \mathcal{C}/\{c_i,c_j\} \cup \{c\}$
	\EndIf
      \EndFor
      \State 
      \Return $\mathcal{C}$
    \EndProcedure
  \end{algorithmic}
\end{algorithm}

\begin{table*}[]
\centering
\caption{Sample Clusters and their Contents in various segments}
\label{ClusterContents}
\begin{tabular}{@{}lll@{}}
\toprule
Segment & Sample Cluster Names                                                                                         & Sample Queries                                                                                                                                                                                                                                                                                                                                                                                 \\ \midrule
Retail  & \begin{tabular}[c]{@{}l@{}}christmas lights\\ shop home\\ gift cards\\ shirts nfl\\ deals cyber\end{tabular} & \begin{tabular}[c]{@{}l@{}}led christmas lights, christmas laser lights,fiber optics christmas trees\\ whirlpool refrigerators, wallpaper, kitchen aid mixer, flannel sheets\\ gift baskets, harry and david, amazon gift card, visa gift card\\ nfl shop, nfl jerseys, champion sportswear, fanatics sports apparel\\ amazon cyber monday, target cyber monday, cyber monday ads\end{tabular} \\ \midrule
Health  & \begin{tabular}[c]{@{}l@{}}heart symptoms\\ health insurance\end{tabular}                                    & \begin{tabular}[c]{@{}l@{}}tachycardia, bradycardia, arrythmia, palpitations, heart rate, pacemaker\\ oscar health insurance, humana health insurance, ehealthinsurance\end{tabular}                                                                                                                                                                                                           \\ \midrule
Travel  & \begin{tabular}[c]{@{}l@{}}las vegas\\ flights cheap\end{tabular}                                            & \begin{tabular}[c]{@{}l@{}}red rock casino, mgm grand, monte carlo las vegas, west gate las vegas\\ expedia, cheaptickets.com flights, cheap airline tickets, travelocity\end{tabular}                                                                                                                                                                                                         \\ \midrule
Phones  & \begin{tabular}[c]{@{}l@{}}microsoft lumia\\ apple iphone\end{tabular}                                       & \begin{tabular}[c]{@{}l@{}}microsoft lumia 650, nokia lumia 640, nokia windows phone, lumia 950xl\\ iphone 6s, iphone 7 release date, iphone 6s case, t mobile iphone\end{tabular}                                                                                                                                                                                                             \\ \bottomrule
\end{tabular}
\end{table*}

In the Merge Phase, distance between pair of clusters is calculated, by computing the distance between their centres and if it is below a threshold value ($MergeThr$), the clusters are merged, and their centres recalculated (In centre recalculation, impression and entropy of the 2 clusters are taken into account, similar to the centre update in Algorithm \ref{DMAlgo}, but not shown in Algorithm \ref{Procedures} to keep it compact). The list of centres ($\mathcal{C}$) is then updated to remove the old centres and the newly merged centre is added to the list.

\section{Results} \label{Results}

We have implemented our approach in Python and also developed a visualization tool using D3.js (\newcite{bostock2011d3}) for viewing the clusters. The tool is being internally used at Microsoft to gain insights on the major areas users searching on Bing are interested in. We have applied our approach to multiple datasets - queries taken from 4 segments - Retail, Health, Travel and Phones.

The statistics of various segments - the number of queries in a segment, the number of clusters formed and the time taken are give in Table \ref{ClusterStats}. The scalability of the approach is evident from the time taken (which does not include time to mine the search logs), where even when the query set is huge, i.e 100 K queries, we have the algorithm completing in just over 5 minutes. The algorithm is highly parallelizable and can be distributed on number of systems in case of even larger datasets, hence making our approach scalable to real-size data that modern search engines receive. 

\begin{figure*}[ht!]
\centering
\includegraphics[width=\textwidth,height=10cm,keepaspectratio]{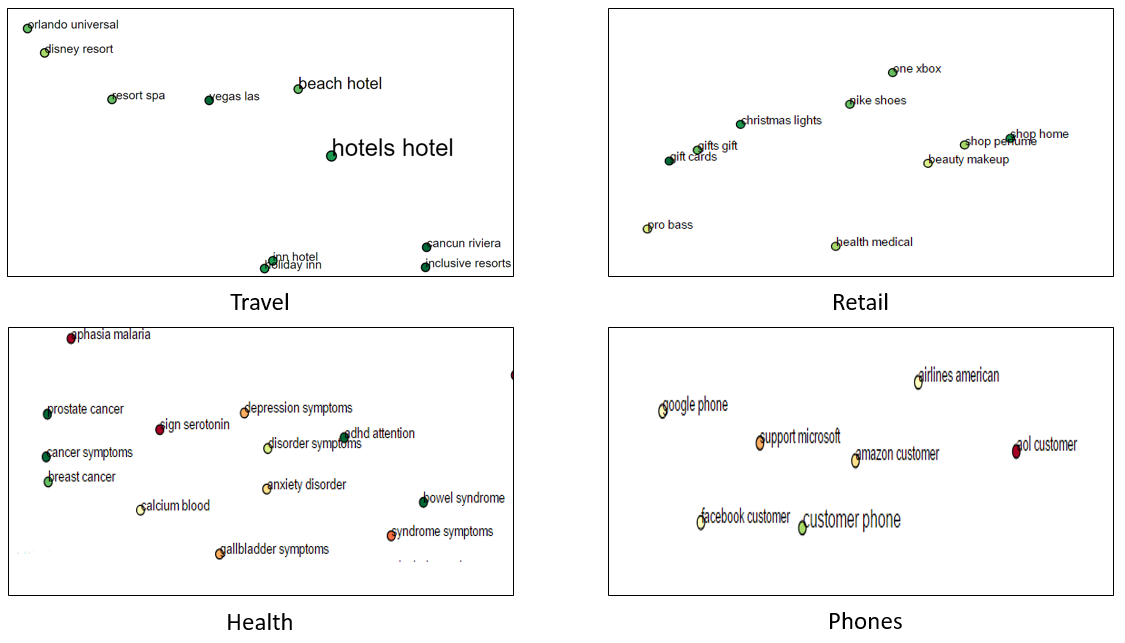}
\caption{TSNE Plots of Clusters in various segments (Small portion displayed) \label{TSNEImages}}
\end{figure*}

\begin{figure}[ht!]
\centering
\includegraphics[width=8cm,height=15cm,keepaspectratio]{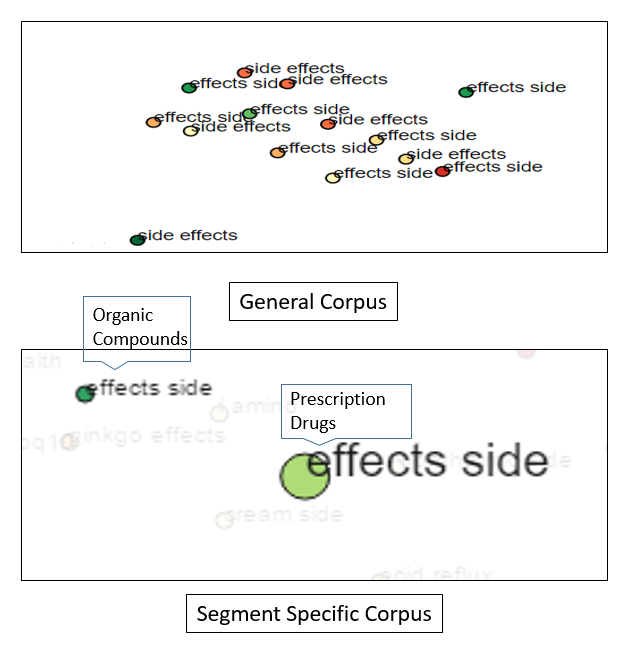}
\caption{Advantage of using Segment Specific Corpus : General Corpus produces a clutter of `side effects' clusters while segment
specific corpus produces 2 `side effects' clusters, each dealing with a different issue\label{SideEffects}}
\end{figure}

In Table \ref{ClusterContents}, we present some sample clusters resulting in each segment and some sample queries within the clusters. The cluster names are derived by finding the 2 most common words in the cluster (considering words from queries and their contexts). It can be observed that the generated clusters are highly meaningful, and the queries within the clusters are also very relevant to the topic of  the cluster. In Figure \ref{TSNEImages}, a small part of the output of visualization tool is shown, for each of the segments. It's a two-layer visualization tool, which contains the clusters in the first layer, and on clicking the cluster, leads to the second layer, which has the contents of the cluster. The higher dimensional vectors (of queries and cluster centres) are plotted in the 2-D space, using the dimensionality reduction technique introduced in \newcite{maaten2008visualizing}, called TSNE. The colours and font-size can be  potentially used to visualize different aspects of a cluster, for example, the revenue earned, user engagement in terms of clicks, etc.

\begin{figure}[ht!]
\centering
\includegraphics[width=8cm,height=15cm,keepaspectratio]{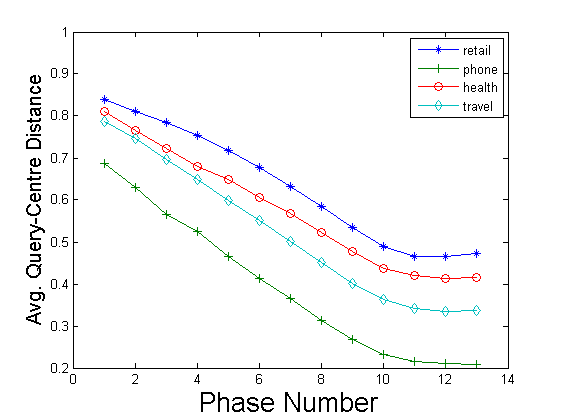}
\caption{Plot of average Query-Centre distance vs Phase Number.\label{AvgQC}}
\end{figure}

In order to measure the effectiveness of the Divide \& Merge Clustering algorithm, we computed the average distance of queries to their centres at each of the divide and merge phases, the plot of which is shown in Figure \ref{AvgQC}. It is evident from the plot of how the average distance decreases with phase, indicating an increase in precision. Since the last is a merge phase, a slight increase in the average distance is observed, but we observed a significant increase in recall. We demonstrate the effectiveness of using segment specific corpus through Figure \ref{SideEffects}. By training the word2vec model on a segment specific corpus (corpus created using the technique from Section \ref{Query2Vec}), we get 2 clusters termed `side effects', with one about organic compounds and the other regarding prescription drugs. But on training with a general corpus (using Google's pretrained vectors \footnote{available at https://code.google.com/archive/p/word2vec/}, trained on Google News Corpora), we get a clutter of `side effects' clusters, with no real significance.

\begin{table}[]
\centering
\caption{Statistics of various segments}
\label{ClusterStats}
\begin{tabular}{@{}llll@{}}
\toprule
Segment & \begin{tabular}[c]{@{}l@{}}Num \\ Queries\end{tabular} & \begin{tabular}[c]{@{}l@{}}Num \\ Clusters\end{tabular} & \begin{tabular}[c]{@{}l@{}}Time taken\\ (in sec)\end{tabular} \\ \midrule
Retail  & 100K                                                   & 271                                                     & 377                                                           \\
Health  & 25K                                                    & 194                                                     & 156                                                           \\
Phones  & 5.7K                                                   & 108                                                     & 41                                                            \\
Travel  & 18.2K                                                  & 164                                                     & 130                                                           \\ \bottomrule
\end{tabular}
\end{table}

\section{Conclusion} \label{Conclusion}

In conclusion, this paper presents a novel unsupervised and scalable algorithm for clustering of search queries, as a means to identify the major interest areas of users. The algorithm uses word2vec in a 2-step fashion to generate vectors for queries and then applies Divide \& Merge Clustering on top of it (which is both computationally efficient and does not require prior knowledge of the number of centres) to generate meaningful clusters.

\bibliographystyle{acl}
\bibliography{Draft2}

\end{document}